# Lorentz-contraction formula from non-simultaneous events

Aleksandar Gjurchinovski

*Institute of Physics, Faculty of Natural Sciences and Mathematics,*
*Sts. Cyril and Methodius University,*
*P. O. Box 162, 1000 Skopje, Macedonia*

E-mail: agjurcin@pmf.ukim.mk

## ABSTRACT

We present a derivation of the relativistic length-contraction formula based on Lorentz space-time transformations on non-simultaneous events. Our derivation avoids the disputable story about the stationary observer and its simultaneous measurements of object's end-points.

**PACS Numbers: 03.30+p**

In the standard introductory textbooks on special relativity, the length of a moving object is defined as the distance between two events that occur simultaneously at the object's end-points [1,2,3]. A simple use of Lorentz transformation would then lead to the conclusion that a uniformly moving object will undergo a length contraction along the direction of its motion.

By strictly considering the length of a uniformly moving object as the distance between its two end-points recorded simultaneously by a stationary observer, the students are sometimes tempted into a misleading interpretation of the contraction effect by overestimating the role played by the observer and the process of relativistic measurement. They consider the length-contraction as inherent to the process of relativistic measurement of the length, therefore being an artificial, apparent phenomenon that does not actually exist in reality [4]. Accordingly, the length of a uniformly moving object will be *measured* to be Lorentz contracted along the direction of its motion, while its *physical length* will remain the same and equal to the object's proper length with respect to every inertial reference frame.

In several recent pedagogical papers, it has been clearly emphasized that Lorentz contraction is a real, physical deformation of a uniformly moving object, a phenomenon that exists regardless of the process of relativistic measurement by the observer [5,6,7]. Adopting the latter approach, in this note we propose a simple derivation of the relativistic contraction formula based on Lorentz space-time transformations of non-simultaneous events. Our derivation, unlike the usual one, clearly emphasizes the physical reality of Lorentz contraction by avoiding the disputable story about the stationary observer and its simultaneous measurements of object's end-points.

Consider a motionless rigid rod oriented parallel to the $X'$ axis (see Fig. 1). Let $A'$ (0,0,0,0) and $B'$ ($l_0$,0,0,$\Delta\tau$) be two space-time events that represent the left and right ends of the rod respectively, such that the event $B'$ will occur at the right end of the rod a time $\Delta\tau$ later than the occurrence of the event $A'$ at the left end of the rod [8]. For example, these two events may be associated with two consecutive explosions of two firecrackers placed at the opposite ends of the rod. The experiment could be arranged in such a manner that the explosion of each of the two firecrackers will leave a sharp mark on the ground that coincides with the position of the rod's end at the time of the explosion. By measuring the distance between the marks on the ground, the observer will measure the length $A'B' = l_0$ of the stationary rod. Obviously, simultaneity of the events $A'$ and $B'$ becomes irrelevant because the rod is stationary. If we use Lorentz transformation equations

$$x = \gamma(x' + vt'), \tag{1}$$

$$y = y', \tag{2}$$

$$z = z', \tag{3}$$

$$t = \gamma(t' + vx'/c^2), \tag{4}$$

to express events $A'$ and $B'$ in $XOY$ frame with respect to which the rod is in uniform rectilinear motion at a constant velocity $v$ along the positive $X$ axis (see Fig. 2), we obtain

$$A(0,0,0,0),\tag{5}$$

and

$$B\left[\gamma(l_0 + v\Delta\tau), 0, 0, \gamma(\Delta\tau + l_0 v/c^2)\right],\tag{6}$$

where $\gamma = 1/\sqrt{1-v^2/c^2}$. From the space-time coordinates of the corresponding events $A$ and $B$ in the reference frame where the rod is moving we see that they will happen at its end-points at different times (see Fig. 2). The event $A$ will happen first, and the event $B$ will happen a time $\gamma(\Delta\tau + l_0 v/c^2)$ later than event $A$. Because the rod is moving, during the time interval $\Delta t = \gamma(\Delta\tau + l_0 v/c^2)$ between the events $A$ and $B$, the rod will traverse the distance $AA'' = B''B = v\Delta t = v\gamma(\Delta\tau + l_0 v/c^2)$ to the right. Let us suppose that the physical length of the moving rod in $XOY$ frame is $AB'' = A''B = l$. Then, according to Fig. 2, we have

$$AB = AB'' + B''B = AA'' + A''B = l + v\gamma(\Delta\tau + l_0 v/c^2)\tag{7}$$

The space separation of the events $A$ and $B$ also follows from (5) and (6):

$$AB = \gamma(l_0 + v\Delta\tau).\tag{8}$$

Thus, by measuring the distance between the marks on the ground made by the explosions from the end-points of the moving rod, the observer in $XOY$ frame will now measure the distance $AB$ in Eqs. (7) and (8). Furthermore, from Eqs. (7) and (8) we obtain

$$l + v\gamma(\Delta\tau + l_0 v/c^2) = \gamma(l_0 + v\Delta\tau),\tag{9}$$

and hence

$$l = \gamma^{-1} l_0 = l_0 \sqrt{1-v^2/c^2},\tag{10}$$

which is the standard Lorentz contraction formula.

In the derivation of Eq. (10) we have used Lorentz transformation formulas on space-time coordinates describing the explosions of the two firecrackers at the rod's ends. The derivation obviously omits the concept of simultaneous measurements of object's end-points. Consequently, it emphasizes the fact that the length of a uniformly moving object does not depend on the process of measurement, and that the object will be physically contracted along its velocity vector by the usual Lorentz factor.

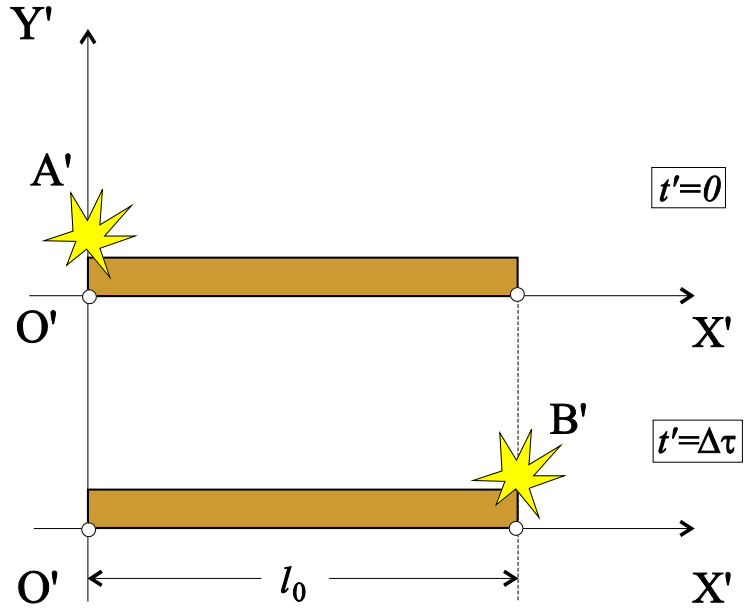

**Fig. 1.** Explosions of the firecrackers from the end-points of the rod in the case when the rod is stationary. The explosion at the left end of the rod (event $A'$) will happen at time $t'=0$, while the explosion at its right end (event $B'$) will happen at time $t'=\Delta\tau$.

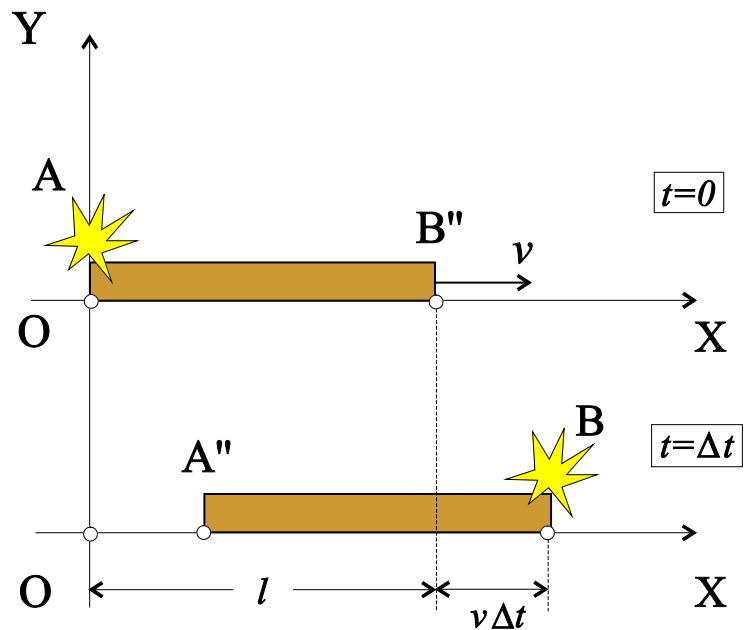

**Fig. 2.** A schematic of the explosions in the case when the rod is moving at a constant velocity $v$ to the right. In this case, the time interval between the corresponding events $A$ and $B$ equals $\Delta t$.